\documentclass[aps,pre,reprint,superscriptaddress]{revtex4-1}

\pdfoutput=1

\usepackage{graphicx}
\usepackage{amssymb}
\usepackage{amsmath}
\usepackage{xcolor}
\usepackage{pdfpages}
\usepackage{pgffor}
\usepackage{CJK}
\usepackage{multirow}
\usepackage{gensymb}
\usepackage{float}

\makeatletter
\AtBeginDocument{\let\LS@rot\@undefined}
\makeatother

\begin{document}

\title{Chain Conformations and Phase Separation in Polymer Solutions with Varying Solvent Quality} 

\author{Yisheng Huang}
\affiliation{Department of Physics, Center for Soft Matter and Biological Physics, and Macromolecules Innovation Institute, Virginia Polytechnic Institute and State University, Blacksburg, Virginia 24061, USA}
\author{Shengfeng Cheng}
\email{chengsf@vt.edu}
\affiliation{Department of Physics, Center for Soft Matter and Biological Physics, and Macromolecules Innovation Institute, Virginia Polytechnic Institute and State University, Blacksburg, Virginia 24061, USA}
\affiliation{Department of Mechanical Engineering, Virginia Polytechnic Institute and State University, Blacksburg, Virginia 24061, USA}

\begin{abstract}
Molecular dynamics simulations are used to investigate the conformations of a single polymer chain, represented by the Kremer-Grest bead-spring model, in a solution with a Lennard-Jones liquid as the solvent when the interaction strength between the polymer and solvent is varied. Results show that when the polymer-solvent interaction is unfavorable, the chain collapses as one would expect in a poor solvent. For more attractive polymer-solvent interactions, the solvent quality improves and the chain is increasingly solvated and exhibits ideal and then swollen conformations. However, as the polymer-solvent interaction strength is increased further to be more than about twice of the strength of the polymer-polymer and solvent-solvent interactions, the chain exhibits an unexpected collapsing behavior. Correspondingly, for strong polymer-solvent attractions, phase separation is observed in the solutions of multiple chains. These results indicate that the solvent becomes effectively poor again with very attractive polymer-solvent interactions. Nonetheless, the mechanism of chain collapsing and phase separation in this limit differs from the case with a poor solvent rendered by unfavorable polymer-solvent interactions. In the latter, the solvent is excluded from the domain of the collapsed chains while in the former, the solvent is still present in the pervaded volume of a collapsed chain or in the polymer-rich domain that phase separates from the pure solvent. In the limit of strong polymer-solvent attractions, the solvent behaves as a glue to stick monomers together, causing a single chain to collapse and multiple chains to aggregate and phase separate.
\end{abstract}

\maketitle

\section{INTRODUCTION}

The study of chain conformations has occupied a central spot in polymer physics as it provides a foundation to understanding the structure and dynamics of polymers.\cite{Wang2017,Kuei2017} For a polymer solution, chain conformations are the outcome of the interactions among the monomers and solvent molecules and thus depend on the solvent quality and polymer concentration. In a dilute solution, chains are not overlapped and the conformation of a single chain is fully determined by the solvent quality and chain length.\cite{RubinsteinColbyBook} As the solvent quality is varied from poor to good, a chain can adopt a collapsed, ideal, or swollen conformation with different scaling behavior of the chain size, often quantified as its radius of gyration, $R_g$, with respect to the molecular weight of the chain. To facilitate the discussion, here we focus on a linear polymer that can be described as a Kuhn chain consisting of $n$ Kuhn bonds, each of length $b$. Then different conformations can be classified by examining the scaling dependence of $R_g$ on $n$ expressed as $R_g \sim n^\alpha$, where $\alpha$ is the Flory exponent. According to the mean-field scaling model, the chain is always ideal if $n$ is smaller than the size of a thermal blob, $g_T = b^6/v^2$, where $v$ is the effective excluded volume of the monomer in the solution. For such ideal chains, $R_g \sim n^{1/2}$ and thus $\alpha = 1/2$. In a $\theta$-solvent, $v= 0 $ and therefore $g_T = \infty$. As a result, all chains in a $\theta$-solvent are ideal. In a poor solvent, a chain with $n > g_T$ has a collapsed conformation with $R_g \sim n^{1/3}$, indicating $\alpha = 1/3$. On the other hand, the chain is swollen in a good solvent with $\alpha = 0.588$.\cite{Guillou1977} In a nonsolvent or an athermal solvent, $g_T = 1$ and the chain is fully collapsed in the former while fully swollen in the latter on the scale of the Kuhn bond.

Experimentally, polymer conformations can be probed with various scattering and spectroscopic techniques.\cite{Hadjichristidis1978,Dondos1996,Armand1998,Grohens1999,Krasovitski2004,Shogbon2006,Essafi2009,Antoniou2010,Traiphol2010,Chen2013,McCulloch2013,Goosen2015,Yakimansky2016,Tenopala-Carmona2018,Zhu2020,Yu2020} Further insight, especially the link between conformations and molecular-scale interactions, can be revealed with the help of molecular modeling methods.\cite{Kremer1988,Kremer1990molecular} Early work mainly used Monte Carlo simulation techniques.\cite{Kremer1988} In the last four decades, molecular dynamics (MD) simulations have played an increasingly important role in understanding polymer conformations and their molecular origin.\cite{Rapaport1978,Rapaport1979,Bishop1979,Bruns1981a,Bruns1981b,Bishop1983,Khalatur1986,Toxvaerd1987,Smit1988ChemPhysLett,Smit1988JCP,Smit1989,Luque1989,Smith1992,Dunweg1993,Grest1993,Grest1994,Kong1997,Luna1997,Ahlrichs1999,Chang2001,Pan2002,Vasilevskaya2003,Polson2002,Polson2005,Steinhauser2005,Dimitrov2007,Tian2009,Zhou2009,Pham2009,Huang2014,Wijesinghe2016,Chremos2018,Wu2018,Lin2021,Dhabal2021}

Rapaport conducted early MD simulations to study relatively short chains that were either isolated \cite{Rapaport1978} or immersed in an explicit solvent \cite{Rapaport1979} based on the hard-sphere model and computed their equilibrium conformational properties such as mean square end-to-end distance and $R_g$. Bishop \textit{et al.} employed MD simulations based on a Lennard-Jones (LJ) 12-6 potential and the finite extensible nonlinear elastic (FENE) bond model to study the static and dynamic properties of a single polymer chain in solution.\cite{Bishop1979} Subsequently, several reports appeared and led to a controversy over the size of an isolated chain compared to that of the same chain immersed in a solvent and cast doubt on the effect of solvents on chain conformations.\cite{Bruns1981a,Bruns1981b,Bishop1983,Khalatur1986,Toxvaerd1987} This issue was addressed by Smit \textit{et al.}, who performed MD simulations for chains based on LJ potentials for nonbonded interactions and a harmonic potential for bonded interactions and clearly demonstrated the effect of solvent quality \cite{Smit1988ChemPhysLett,Smit1989} and density \cite{Smit1988JCP} on chain conformations. Subsequently, Luque \textit{et al.} compared the FENE bond with the harmonic bond and studied the difference caused by different bond models in the static and dynamic properties of polymer chains in solution.\cite{Luque1989} Smith and Rapaport used MD models with purely repulsive LJ potentials for nonbonded interactions and a LJ-based bond potential to investigate the structural and relaxation properties of a linear chain in a solvent.\cite{Smith1992}

In the following decades, MD modeling has evolved into a powerful tool for studying polymer conformations and dynamics.\cite{Kremer1990,Kremer1990molecular,Dunweg1991,Dunweg1993} Particularly, the effects of solvent quality on the static and dynamic properties of polymers have been studied with MD simulations in a wide range of settings.\cite{Grest1993,Grest1994,Kong1997,Luna1997,Chang2001,Pan2002,Vasilevskaya2003,Polson2002,Polson2005,Steinhauser2005,Dimitrov2007,Zhou2009,Wijesinghe2016,Chremos2018,Wu2018} One underlying reason is that it is relatively straightforward to tune the quality of a solvent in a MD model. For example, the relative importance of the monomer-monomer attraction can be varied by changing temperature \cite{Grest1993,Grest1994,Graessley1999} or the depth of the potential governing this interaction \cite{Steinhauser2005} and as a result, the effective quality of the solvent changes without explicitly including solvent atoms. In a supercritical fluid, the solvent quality can also be varied by controlling the relative location of its thermodynamic state with respect to the phase boundary through tuning its density and temperature.\cite{Luna1997} Different solvents of different qualities can be used.\cite{Wijesinghe2016} More frequently, the quality of a solvent can be varied by changing the strength (e.g., the depth of the attractive potential) of the monomer-solvent interaction with respect to those of the monomer-monomer and solvent-solvent interactions.\cite{Smit1988ChemPhysLett,Smit1988JCP,Smit1989,Kong1997,Chang2001,Pan2002,Polson2002, Polson2002,Dimitrov2007,Zhou2009,Chremos2018,Wu2018}

The past MD simulations have provided results that largely corroborate the mean-field predictions of chain conformations in solvents with variable qualities, including the coil-globule transition when the solvent quality is varied.\cite{Polson2005} To reduce the computational cost, purely repulsive interactions among monomers and solvent atoms are often adopted in these simulations. For example, Chang and Yethiraj used purely repulsive LJ 12-6 potentials for all nonbonded interactions to realize a good solvent and then when the monomer-monomer and solvent-solvent interactions were extended to include an attractive tail, the solvent became poor.\cite{Chang2001} In the works of Zhou and Daivis \cite{Zhou2009} and Dimitrov \text{et al.},\cite{Dimitrov2007} purely repulsive LJ 12-6 potentials were used for all nonbonded interactions and when the monomer-solvent interaction was made more repulsive, the solvent quality was reduced. Polson and coworkers used a LJ 12-6 potential with an attractive tail (at a cutoff of 2.5$\sigma$) for the solvent-solvent and monomer-monomer interactions. When the same LJ potential was used for the monomer-solvent interaction, the solvent behaved as a good one. A poor solvent was realized by making the monomer-solvent interaction purely repulsive.\cite{Polson2002,Polson2005} Pure repulsion was also used in the model of Wu \textit{et al.} to render a good excluded volume between certain beads,\cite{Wu2018} where LJ 12-10 potentials were adopted. However, it is unclear if there is any abrupt change brought into these systems when a potential is varied from having an attractive tail to being purely repulsive.

In this paper, our goal is to perform a systematic study to elucidate the solvent effect on chain conformations and the variation of the corresponding Flory exponents as the quality of the solvent is gradually varied. In the model adopted here, LJ potentials with an attractive tail are used for all the nonbonded interactions, with the solvent-solvent and nonbonded monomer-monomer interactions being identical and having a fixed strength. The depth of the monomer-solvent interaction potential is then gradually varied from shallow to deep, targeting solvents with qualities varying from poor to good. A similar approach was used in the early work of Smit \textit{et al.}.\cite{Smit1988ChemPhysLett,Smit1989} Here we extend the range of the monomer-solvent potential depth being probed and show that an unexpected chain collapse occurs when the monomer-solvent attraction is made more than about twice strong as the solvent-solvent and nonbonded monomer-monomer interactions. That is, the solvent becomes effectively poor again with strong monomer-solvent attractions. In this limit, however, the corresponding polymer solutions exhibit a phase separation behavior that differs from the case of a poor solvent rendered by weak monomer-solvent attractions.

\section{SIMULATION METHODS}

The solvent is modeled as point particles interacting through a standard LJ 12-6 potential
\begin{eqnarray}
U_\text{LJ}(r) & = &  4 \epsilon\left[\left(\frac{\sigma}{r}\right)^{12}-\left(\frac{\sigma}{r}\right)^{6} \right. \nonumber \\
& & \left. -\left(\frac{\sigma}{r_c}\right)^{12}+\left(\frac{\sigma}{r_c}\right)^{6}\right] \quad \text{for} \quad r\le r_c~,
\label{eq:LJ}
\end{eqnarray}
where $r$ is the distance between two particles, $\epsilon$ is an energy scale dictating the interaction strength, $\sigma$ is a length scale, and $r_c$ is the cutoff of the potential. All the physical quantities will be reported in terms of $\epsilon$, $\sigma$, and the mass of a solvent bead, $m$.

A polymer chain consists of $N_m$ beads of mass $m$ connected by bonds described by a standard FENE potential\cite{Warner1972,Armstrong1974,Bishop1979,Kremer1990,Kremer1990molecular}
\begin{eqnarray}
U_\text{B}(r) & = & -\frac{K}{2}R_{0}^{2} \ln \left[1-\left(\frac{r}{R_{0}}\right)^{2}\right] \nonumber \\
& & +4 \epsilon\left[\left(\frac{\sigma}{r}\right)^{12}-\left(\frac{\sigma}{r}\right)^{6} + \frac{1}{4} \right]~,
\label{eq:FENE}
\end{eqnarray}
where $K= 30 \epsilon/\sigma^2$ and $R_0 = 1.5\sigma$.\cite{Kremer1990,Kremer1990molecular} In the FENE potential, the first term extends to $R_0$ and the second term is cut off at $2^{1/6}\sigma$. Two monomer beads that are not directly bonded interact through a LJ 12-6 potential in Eq.~(\ref{eq:LJ}) with an interaction strength $\epsilon_{pp} = \epsilon$. The monomer-solvent interaction is given by an additional LJ 12-6 potential with an interaction strength $\epsilon_{ps} = \lambda\epsilon$ with $\lambda$ varying from 0.4 to 4.0. The cutoff of all the LJ interactions for the solvent-solvent, solvent-monomer, and nonbonded monomer-monomer pairs is set at $r_c = 3.0\sigma$ and the LJ potentials are shifted to 0 at $r=r_c$. Smit \textit{et al.} used a similar model to study the influence of solvent quality on polymer properties, where $r_c$ was set at 2.5$\sigma$ and $\lambda$ was varied from 0 to 1.4.\cite{Smit1988ChemPhysLett,Smit1989}

All the simulations are performed with the Large-scale Atomic/Molecular Massively Parallel Simulator (LAMMPS).\cite{Plimpton1995} The equation of motion is integrated with a velocity Verlet algorithm \cite{Verlet1967} with a time step $\Delta t= 0.005\tau$, where $\tau = \sqrt{m\sigma^2/\epsilon}$ is the LJ unit of time. A cubic simulation box is employed with periodic boundary conditions applied to all directions. All simulations reported here are performed at a constant temperature $T = 1.0\epsilon/k_\text{B}$, where $k_\text{B}$ is the Boltzmann constant, and a constant pressure $P = 0.05 \epsilon/\sigma^3$. The temperature is controlled by a Nose-Hoover thermostat with a damping time $\Gamma = 100 \tau$ and pressure control is realized with a Nose-Hoover barostat.\cite{Shinoda2004} The pressure is controlled in a hydrostatic manner such that the simulation box remains cubic in all runs.

\section{RESULTS}

\subsection{Single-chain conformations}

\begin{figure}[htb]
   \centering
   \includegraphics[width=0.35\textwidth]{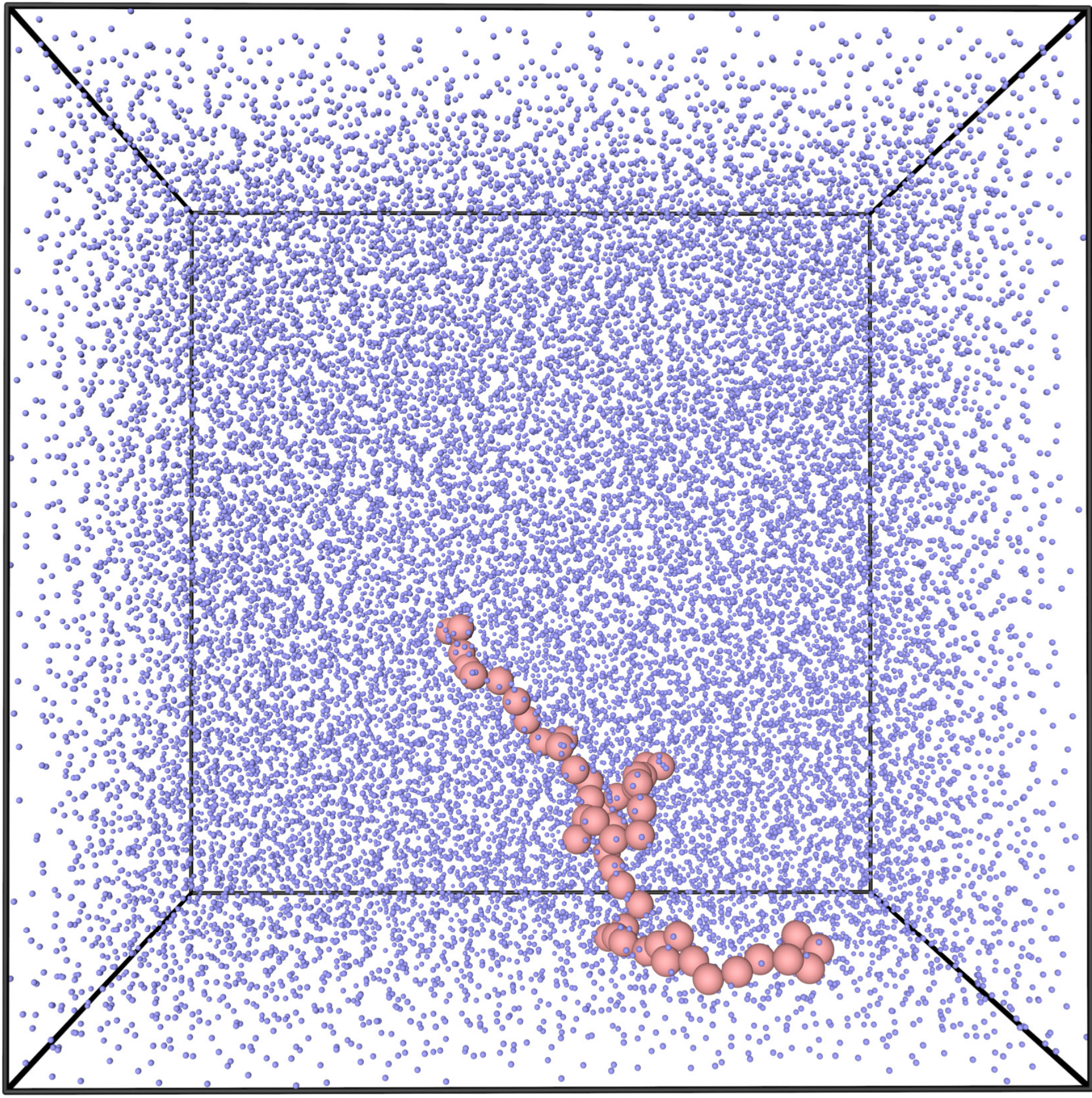}
   \caption{Snapshot of a 50-bead chain suspended in the LJ solvent at $\epsilon_{ps} = 2.0\epsilon$.}
 \label{fig:single_chain_sys}
\end{figure}

Figure \ref{fig:single_chain_sys} shows a typical system where a single chain of 50 beads is suspended in the LJ solvent. At a given $\epsilon_{ps}$, linear chains of different lengths with $N_m$ from $16$ to $512$ are simulated. The number of bonds is $N=N_m-1$. Each system is equilibrated for at least $10^4\tau$ to allow the chain to fully relax. In the equilibrium state, the solvent density is about 0.64$m/\sigma^3$.\cite{Cheng2013JCP} In the following production run, which is at least $2.5\times 10^4\tau$, data are collected on chain conformations and structures. For each system, the number of solvent atoms is adjusted such that the size of the cubic simulation box is at least four times larger than the radius of gyration of the chain.

\begin{figure}[htb]
   \centering
   \includegraphics[width=0.4\textwidth]{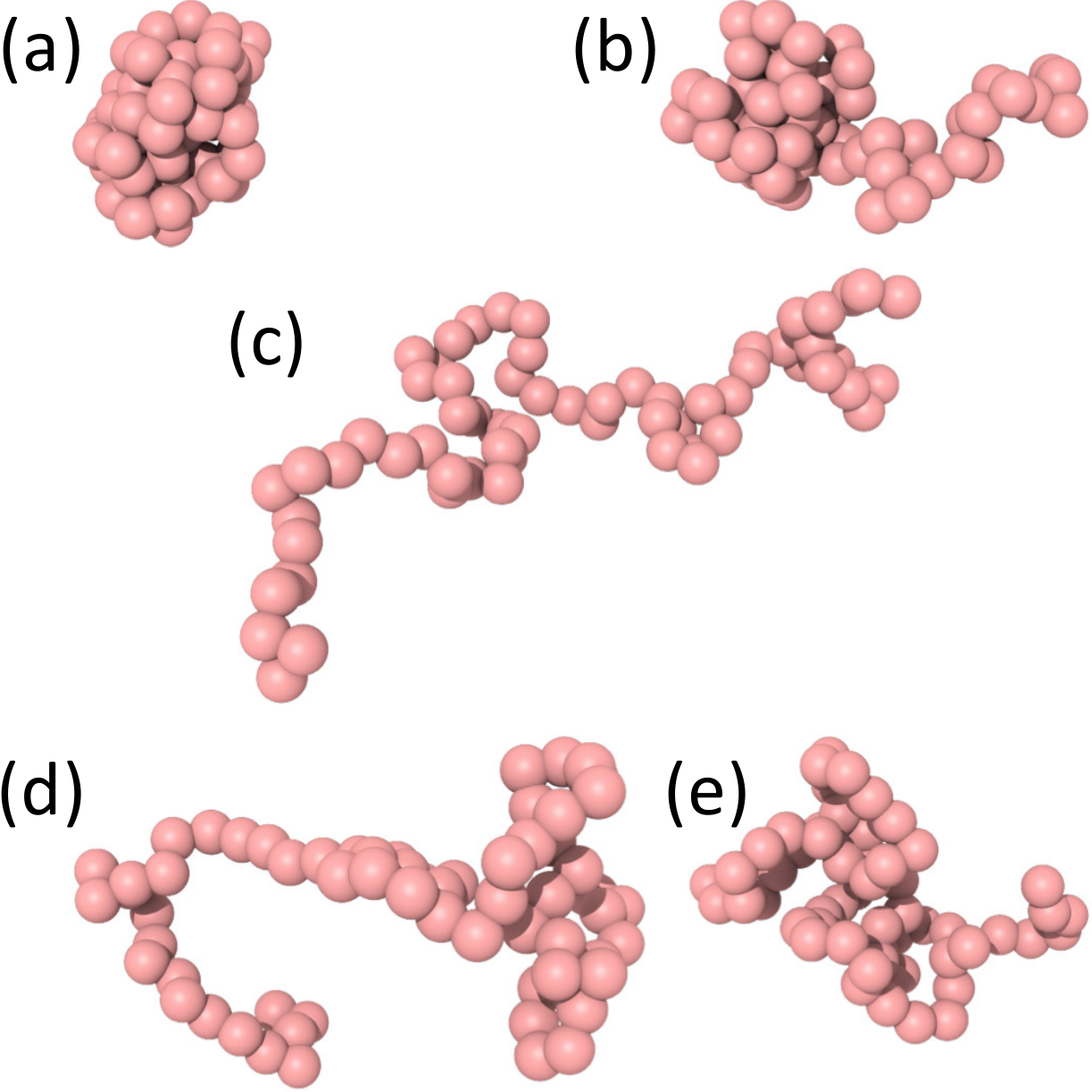}
   \caption{Representative snapshots of a 50-bead chain at various values of $\epsilon_{ps}$: (a) $0.4\epsilon$, (b) $0.95\epsilon$, (c) $1.5\epsilon$, (d) $2.0\epsilon$, and (e) $4.0\epsilon$.}
 \label{fig:single_chain_snap}
\end{figure}

Figure \ref{fig:single_chain_snap} shows the representative snapshots of a 50-bead chain at $\epsilon_{ps}=0.4\epsilon$, $0.95\epsilon$, $1.5\epsilon$, $2.0\epsilon$, and $4.0\epsilon$. As expected, the chain shows a collapsed conformation at small values of $\epsilon_{ps}$ (e.g., Fig.~\ref{fig:single_chain_snap}(a) for $\epsilon_{ps}=0.4\epsilon$) as the solvent quality is rather poor when the monomer-solvent interaction is very unfavorable. As $\epsilon_{ps}$ is increased, the solvent quality improves and chains shorter than a thermal blob exhibit ideal chain conformations. At $\epsilon_{ps}=0.95\epsilon$, all chains studied here seem to be almost ideal and one example is shown in Fig.~\ref{fig:single_chain_snap}(b). When $\epsilon_{ps}$ is increased further, chains adopt swollen conformations. The case in Fig.~\ref{fig:single_chain_snap}(c) for $\epsilon_{ps}=1.5\epsilon$ is one such example.

A surprise is revealed in the simulations with large values of $\epsilon_{ps}$. As shown in Fig.~\ref{fig:single_chain_snap}(d), the chain becomes less extended at $\epsilon_{ps} = 2.0\epsilon$ compared with the case at $\epsilon_{ps} = 1.5\epsilon$. The trend is clearer when $\epsilon_{ps}$ is increased further. For example, the snapshot in Fig.~\ref{fig:single_chain_snap}(e) is for $\epsilon_{ps} = 4.0\epsilon$, where the chain is apparently collapsed. This result is unexpected as a large value of $\epsilon_{ps}$ indicates that the monomer beads interact strongly with the solvent atoms, where we would naively expect the polymer chain to be well solvated by the solvent and therefore to adopt extended conformations. Below we further quantify the variation of chain sizes as $\epsilon_{ps}$ is increased and then discuss the implication of our results.

\begin{figure}[htb]
   \centering
   \includegraphics[width=0.4\textwidth]{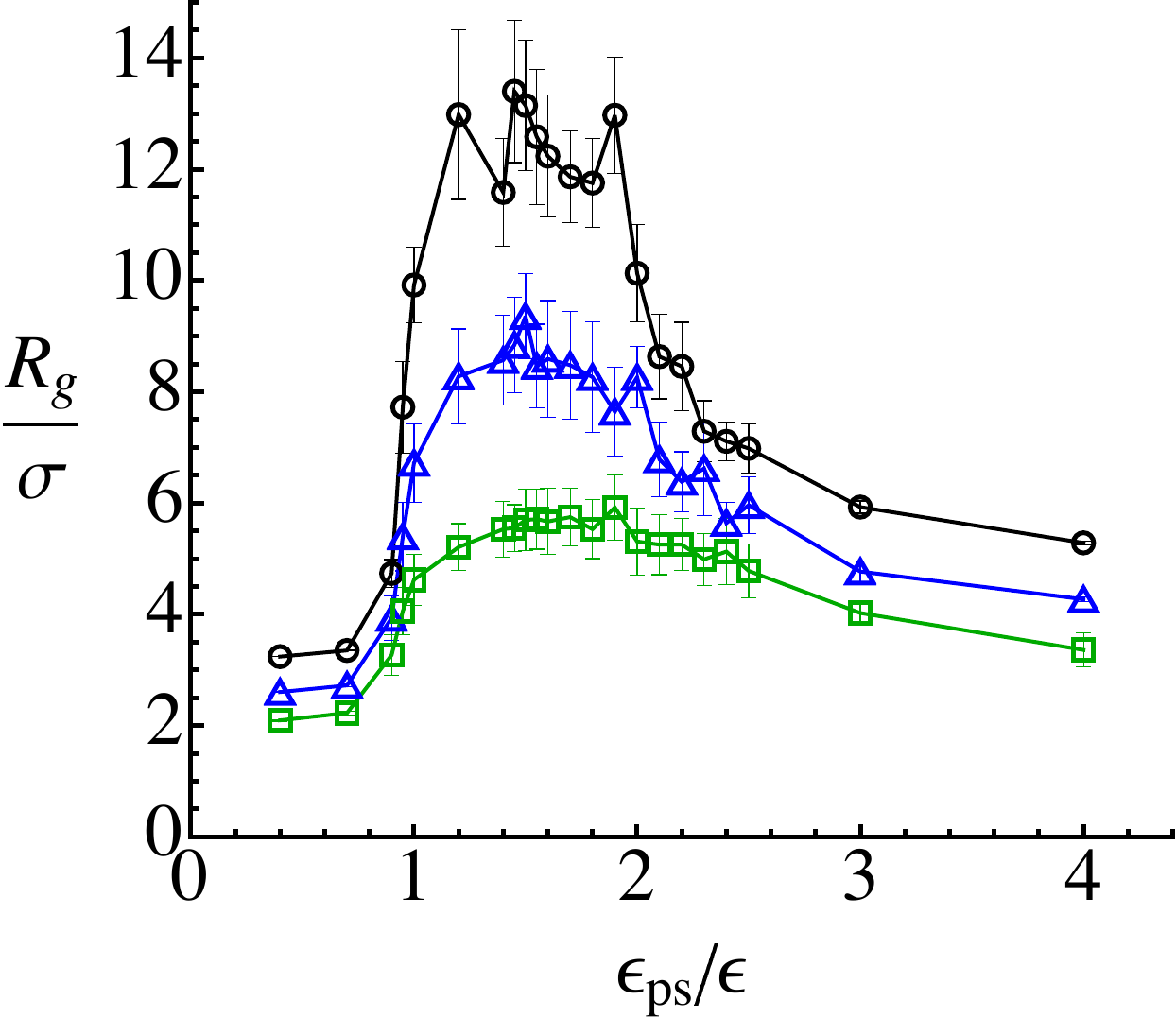}
   \caption{Radius of gyration ($R_g$) vs. monomer-solvent interaction strength ($\epsilon_{ps}$) for $N_m=64$ (squares), $128$ (triangles), and $256$ (circles).}
 \label{fig:Rg_vs_Eps_3N}
\end{figure}

We compute the radius of gyration, $R_g$, of each polymer chain as a way to quantify chain conformations. Since the correlation time of $R_g$ is found to be around or shorter than $2.5\tau$, $R_g$ is computed every 2.5$\tau$ and its average value is output every 50$\tau$. A statistical analysis is then performed for a sequence of such average values to obtain the mean value and uncertainty of $R_g$ reported here. In Fig.~\ref{fig:Rg_vs_Eps_3N}, $R_g$ is plotted against the monomer-solvent interaction strength, $\epsilon_{ps}$, for three chains with $N_m=64$, $128$, and $256$, respectively. Results for other values of $N_m$ studied are all included in the Supporting Information. Consistent with the snapshots shown in Fig.~\ref{fig:single_chain_snap}, $R_g$ is initially small when $\epsilon_{ps}$ is small. A first sharp transition of $R_g$ occurs at $\epsilon_{ps} \simeq 1.0\epsilon$, around which point the solvent quality changes from poor to $\theta$ and then to good. When $\epsilon_{ps}$ is increased beyond about $2.0\epsilon$ (i.e., the monomer-solvent interaction is twice as strong as the solvent-solvent and nonbonded monomer-monomer interactions), $R_g$ starts to decrease, indicating the collapsing of the chain and the worsening of the solvent quality. Therefore, there is another $\theta$-point around $\epsilon_{ps}\simeq 2.0\epsilon$. This second $\theta$-transition is sharper for a longer chain.

\begin{figure}[htb]
   \centering
   \includegraphics[width=0.45\textwidth]{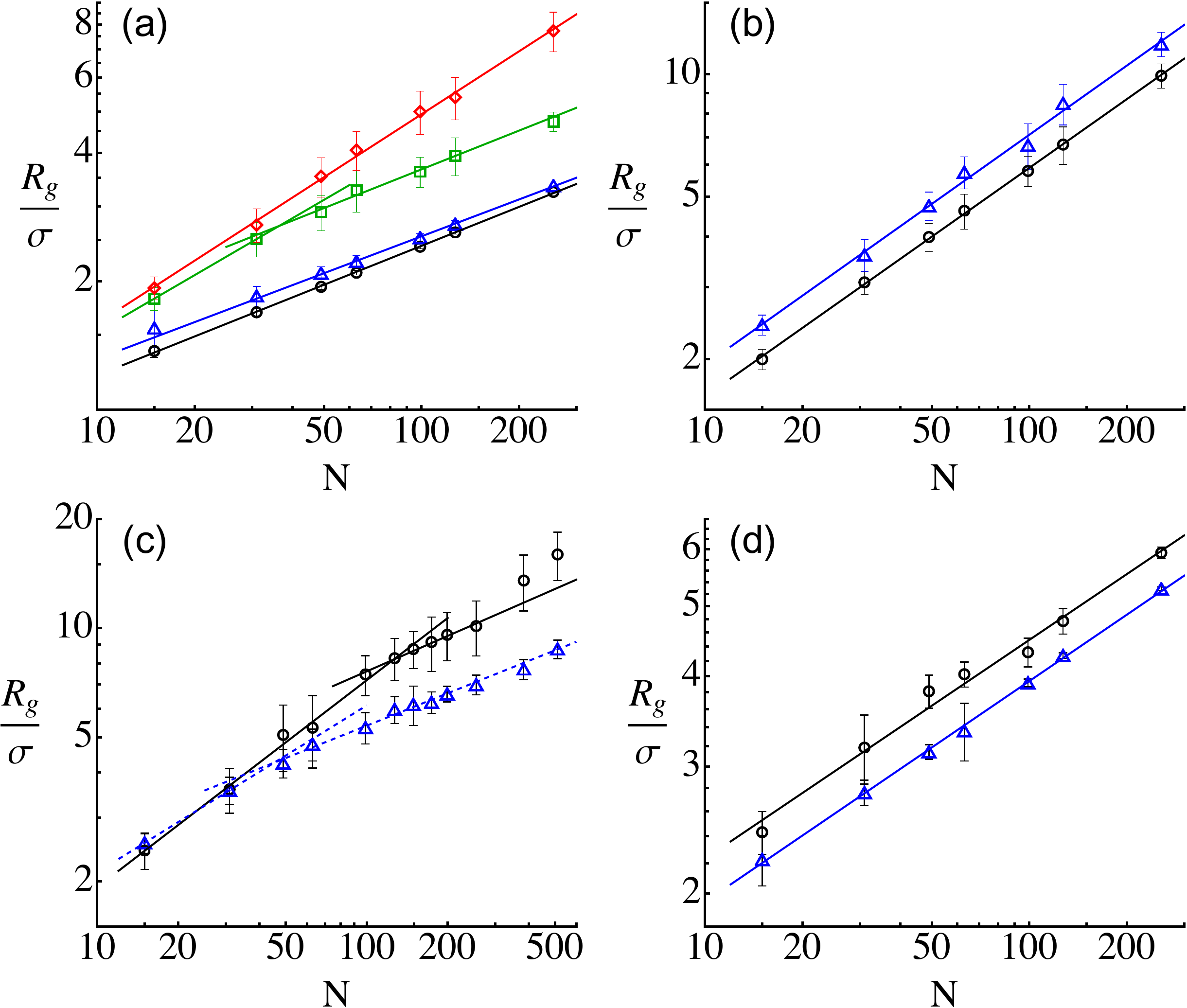}
   \caption{Radius of gyration ($R_g$) vs. number of bonds ($N$) at various values of $\epsilon_{ps}$: (a) $0.4\epsilon$ (circles), $0.7\epsilon$ (triangles), $0.9\epsilon$ (squares), and $0.95\epsilon$ (diamonds); (b) $1.0\epsilon$ (circles) and $1.7\epsilon$ (triangles); (c) $2.0\epsilon$ (circles) and $2.5\epsilon$ (triangles); (d) $3.0\epsilon$ (circles) and $4.0\epsilon$ (triangles).}
 \label{fig:Rg_vs_N_select}
\end{figure}

The variation of chain sizes can also be analyzed by examining the dependence of $R_g$ on the chain length (i.e., the number of bonds), $N$. The representative results are shown in Fig.~\ref{fig:Rg_vs_N_select} and all of the rest are included in the Supporting Information. According to the scaling model,\cite{RubinsteinColbyBook} $R_g \sim N^\alpha$ where $\alpha$ is the Flory exponent. In a $\theta$-solvent or when $N<g_T$ in a good or poor solvent, $\alpha = 1/2$, indicating an ideal-chain behavior. For a chain longer than $g_T$, the renormalization-group result is that $\alpha = 0.588$ in a good solvent \cite{Guillou1977} and the mean-field prediction is that $\alpha = 1/3$ in a poor one.\cite{RubinsteinColbyBook} In an athermal solvent or a nonsolvent, $g_T = 1$ and $\alpha$ is $0.588$ for all chains in the former and $1/3$ for the latter. The results Fig.~\ref{fig:Rg_vs_N_select} can be understood on the basis of the Flory exponent, as detailed below.

As shown in Fig.~\ref{fig:Rg_vs_N_select}(a), for $\epsilon_{ps} =0.4\epsilon$ and $0.7\epsilon$, the values of $\alpha$ are found to be 0.31 and 0.29, respectively, indicating nonsolvent situations. When $\epsilon_{ps}$ is increased to $0.9\epsilon$, $\alpha$ is $\sim 0.44$ for short chains while $\sim 0.30$ for longer chains. This implies that the solvent is still poor at $\epsilon_{ps}= 0.9 \epsilon$, where $g_T$ is around 40. When $\epsilon_{ps}$ is slightly increased further to $0.95\epsilon$, the $\theta$-solvent case seems to be realized and $\alpha$ is about 0.49. The data in Fig.~\ref{fig:Rg_vs_N_select}(b) are for $\epsilon_{ps} =1.0\epsilon$ and $1.7\epsilon$. Both are consistent with a good-solvent behavior with a small $g_T$, reflected by the corresponding values of $\alpha$ being about 0.56 in both systems. Furthermore, at a given $N$ the chain is slightly more swollen for $\epsilon_{ps} = 1.7\epsilon$ than for $\epsilon_{ps} = 1.0\epsilon$.

The snapshots shown in Fig.~\ref{fig:single_chain_snap} indicate that after swelling with $\epsilon_{ps}$ increased to somewhere between $1.5\epsilon$ and $2.0\epsilon$, chain conformations start to contract again once $\epsilon_{ps}$ is increased further. For $\epsilon_{ps} =1.8\epsilon$ and $1.9\epsilon$, the data are included in the Supporting Information and can be fitted to $\alpha \simeq 0.53$ and $0.54$, respectively, revealing a decreasing trend of the Flory exponent. At $\epsilon_{ps} = 2.0\epsilon$, the data on $R_g$ vs. $N$ show two distinct scaling regimes, as shown in Fig.~\ref{fig:Rg_vs_N_select}(c). For small values of $N$, the Flory exponent $\alpha$ is about $0.57$ while $\alpha \simeq 0.33$ when $N$ is large, identifying the solvent as a poor one. A similar trend is seen for $\epsilon_{ps} = 2.5\epsilon$, where $\alpha \simeq 0.46$ for short and $0.30$ for relatively long chains. Furthermore, it can be noted in Fig.~\ref{fig:Rg_vs_N_select}(c) that while shorter chains have similar sizes, for a given long chain $R_g$ adopts a larger value at $\epsilon_{ps} = 2.0\epsilon$ than at $2.5\epsilon$. The implication is that the solvent quality gradually deteriorates as $\epsilon_{ps}$ is increased beyond $2.0\epsilon$. This trend continues to large values of $\epsilon_{ps}$ (i.e., increasing monomer-solvent attractions), including the data shown in Fig.~\ref{fig:Rg_vs_N_select}(d) for $\epsilon_{ps} = 3.0\epsilon$ and $4.0\epsilon$, where $R_g$ is smaller for the latter and the Flory exponent is about 0.30 and 0.31, respectively, close to the nonsolvent result.

\begin{figure}[htb]
   \centering
   \includegraphics[width=0.45\textwidth]{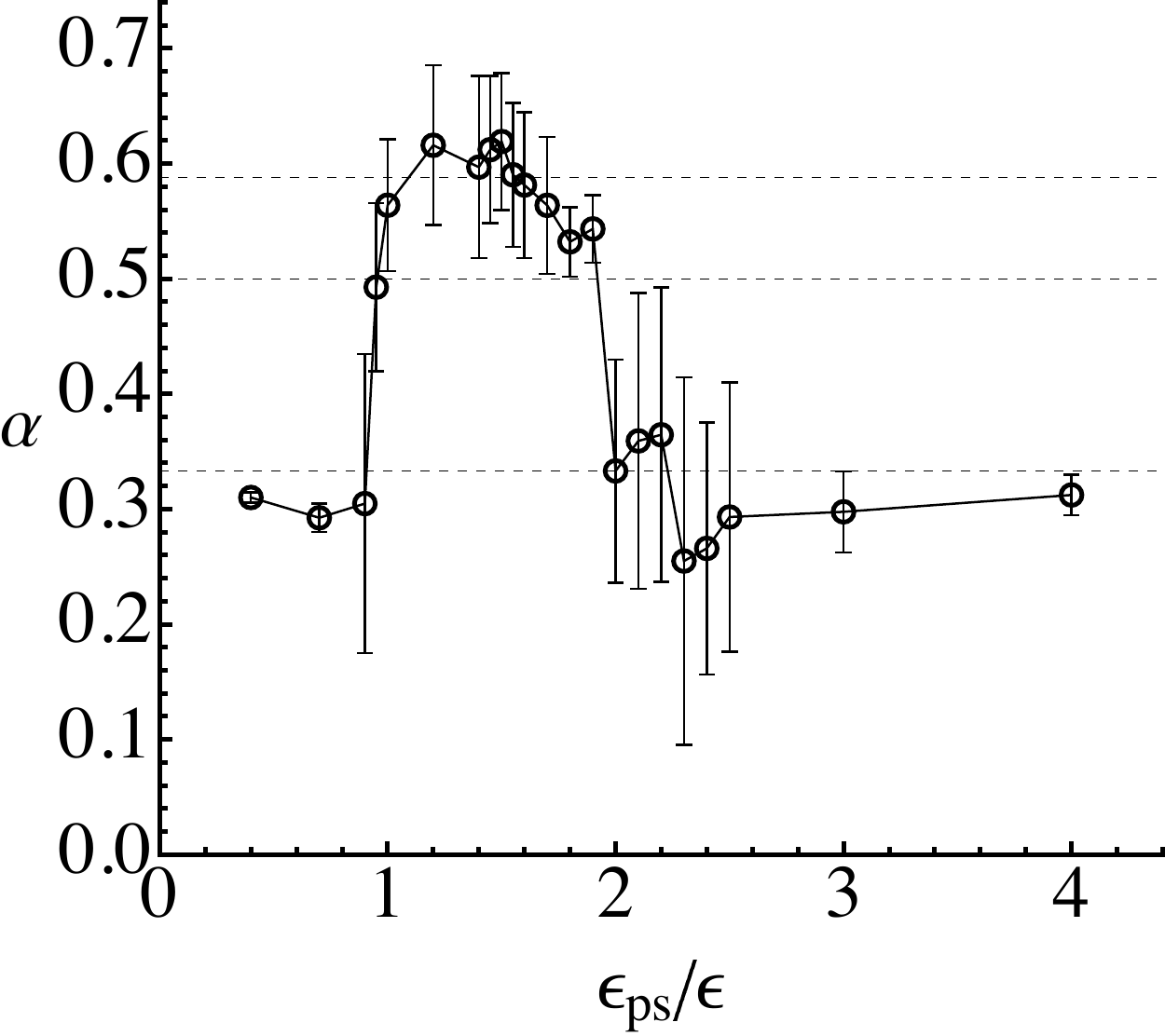}
   \caption{Flory exponent ($\alpha$) vs. monomer-solvent interaction strength ($\epsilon_{ps}$). The dashed lines from top to bottom indicate $\alpha = 0.588$, 0.5, and 1/3, respectively.}
 \label{fig:flory_exponent}
\end{figure}

All the results on the Flory exponent, $\alpha$, are summarized in Fig.~\ref{fig:flory_exponent} for sufficiently long chains (i.e., longer than the thermal blob size). For small values of $\epsilon_{ps}$, $\alpha$ is close to about 0.3, smaller than the mean-field value of 1/3 but very close to the result obtained by Gan and Eu using a statistical mechanical theory based on integral equations derived from the polymer Kirkwood hierarchy.\cite{Gan1998} The first $\theta$-point occurs at $\epsilon_{ps} \simeq 0.95\epsilon$, at which $\alpha \simeq 1/2$. When $\epsilon_{ps}$ is increased beyond this value, $\alpha$ is close to 0.588, the Flory exponent expected for a long chain in a good solvent.\cite{Guillou1977} However, a second $\theta$-point occurs for $\epsilon_{ps}$ somewhere between  $1.9\epsilon$ and $2.0\epsilon$, where $\alpha \simeq 1/2$ again. For even larger values of $\epsilon_{ps}$, $\alpha$ decreases back to about 0.3, indicating another poor-solvent regime. Again, the scaling exponent in this limit of strong monomer-solvent attractions is smaller than the mean field value and closer to the result of Gan and Eu.\cite{Gan1998} It should also be pointed out that for good solvents, the calculations of Gan and Eu using integral equations yield $\alpha  =0.61$. Our MD results on $\alpha$ for the systems with $1.2\epsilon \lesssim \epsilon_{ps} \lesssim 1.5\epsilon$ are closer to this prediction \cite{Gan1998} than the result from the renormalization group analysis of field theory.\cite{Guillou1977}

\begin{figure*}[htb]
   \centering
   \includegraphics[width=\textwidth]{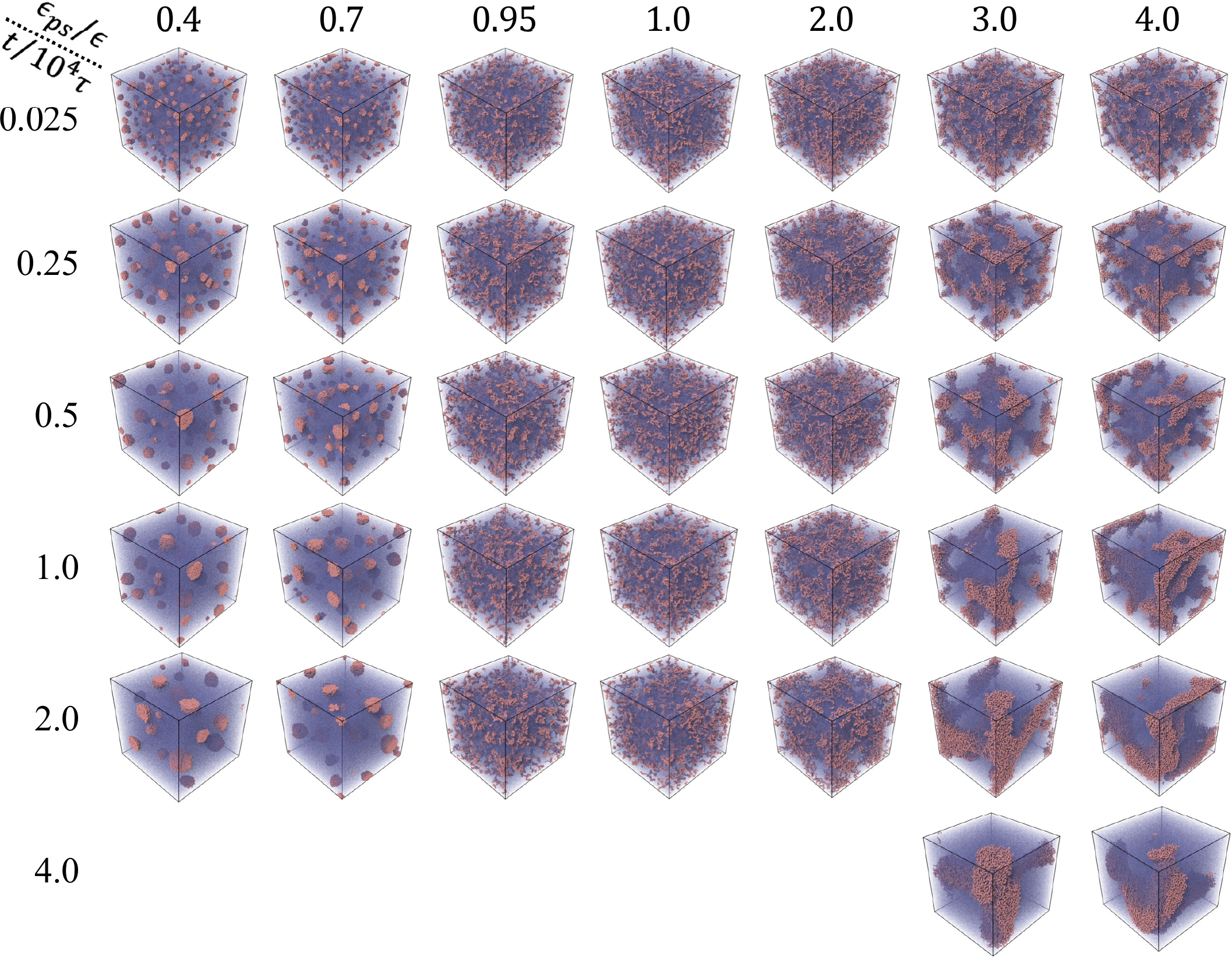}
   \caption{Time evolution of solution morphology at various values of $\epsilon_{ps}$.}
 \label{fig:solution_morphology}
\end{figure*}

\subsection{Polymer solutions}

The single-chain behavior at various monomer-solvent interaction strengths discussed in the previous section has a clear bearing on the morphology of polymer solutions. The time evolution of solution morphology at various values of $\epsilon_{ps}$ is shown in Fig.~\ref{fig:solution_morphology}. In these simulations, 1000 chains of the same length of $N_m=50$ are added to a solvent consisting of 950,000 atoms. The simulations with 500 chains of $N_m=100$ yield very similar results. For both chain lengths, the average number density of monomer beads at different $\epsilon_{ps}$ ranges from $0.032$ to $0.035\sigma^{-3}$, which for $N_m = 50$ ($100$) is about three (two) times smaller than the critical density, about $0.1\sigma^{-3}$ ($0.07\sigma^{-3}$), estimated from $R_g$ at which the pervaded volumes of chains in a good solvent start to overlap. Therefore, all the solutions reported here are in the dilute regime. Below we mainly present results for $N_m=50$. All the solutions start in a state where the polymer chains are well dispersed in the solvent. Simulations are then conducted in a NPT ensemble with $P=0.05\epsilon/\sigma^3$, where the equilibrium solvent density is about 0.64$m/\sigma^3$.\cite{Cheng2013JCP}

The solvent is very poor at $\epsilon_{ps}=0.4\epsilon$ and $0.7\epsilon$, where the polymer chains phase separate from the solvent and form globules. As time passes, the globules grow by adsorbing more chains or merging with each other. The final state is determined by a thermodynamic balance of entropy, which favors more globules dispersed in the solution, and the interfacial tension of the globule-solvent interface, which drives the coalescence and aggregation of globules.

At $\epsilon_{ps}=0.95\epsilon$ and $1.0\epsilon$, the polymer chains stay dispersed in the solvent throughout the simulation, indicating either a $\theta$ or good solution. At $\epsilon_{ps}=2.0\epsilon$, the chains also appear to be dispersed but some local aggregation of chains can be observed, especially in the late stage of the simulation such as the state at $t=2\times 10^4 \tau$. If longer chains are simulated at longer times, a stronger phase separation trend is expected. The situation is quite different for larger values of $\epsilon_{ps}$, such as the cases with $\epsilon_{ps}=3.0\epsilon$ and $4.0 \epsilon$ shown in Fig.~\ref{fig:solution_morphology}. Again, monomer-solvent phase separation is clearly observed. However, this type of phase separation is different from that in solutions with small values of $\epsilon_{ps}$. In the limit of strong monomer-solvent attractions, the polymer chains do not form globules. Instead, they form extended network-like structures percolating the simulation box.

The various types of solution morphology shown in Fig.~\ref{fig:solution_morphology} for different values of $\epsilon_{ps}$ are all consistent with the single-chain conformations discussed early. The solvent is found to be poor at $\epsilon_{ps} \lesssim 0.95\epsilon$ and effectively poor at $\epsilon_{ps} \gtrsim 2.0\epsilon$. The good-solvent regime is only observed for the intermediate range of $\epsilon_{ps}$ that are separated from the two poor-solvent regimes by two $\theta$-transitions. Below we examine the distribution of monomer beads and the correlation between monomers and solvent atoms in more details and aim to reveal a physical picture underlying the poor-solvent phenomenon at large values of $\epsilon_{ps}$.

\begin{figure}[htb]
   \centering
   \includegraphics[width=0.35\textwidth]{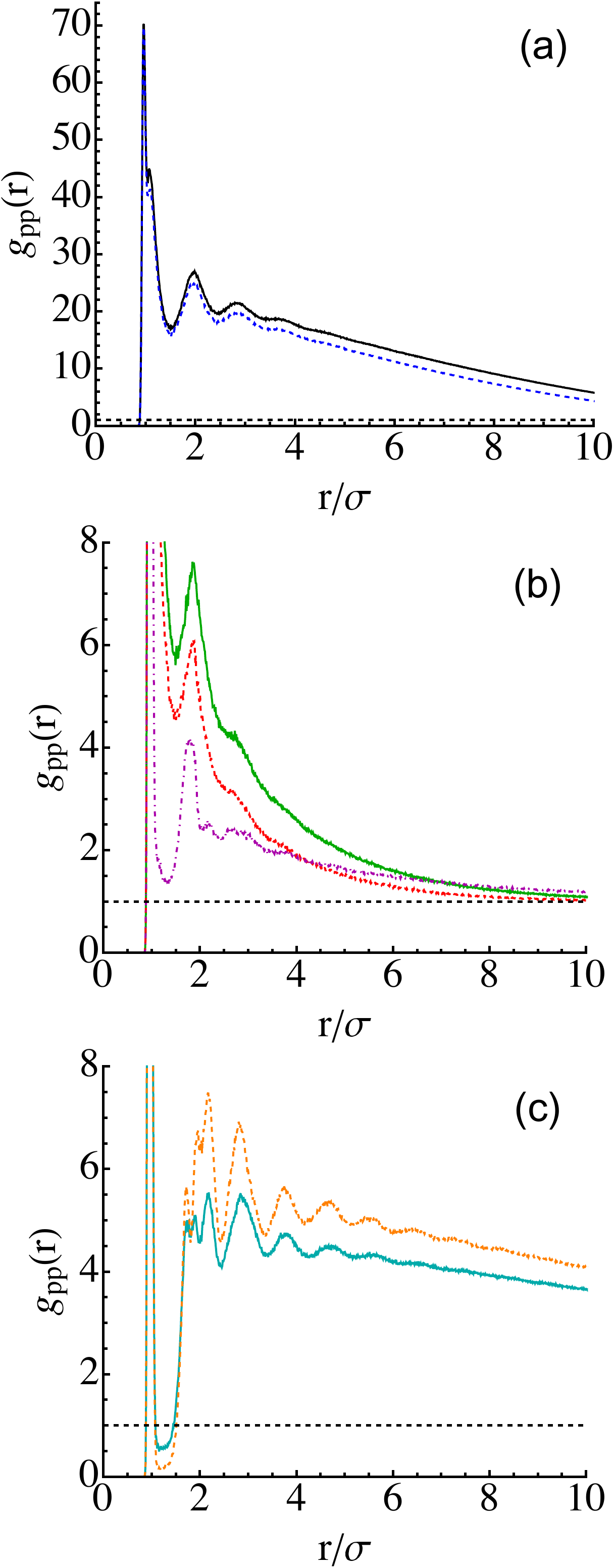}
   \caption{Monomer-monomer radial distribution function, $g_{pp}(r)$, at various values of $\epsilon_{ps}$: (a) $0.4\epsilon$ (black solid line) and $0.7\epsilon$ (blue dashed line); (b) $0.95\epsilon$ (green solid line), $1.0\epsilon$ (red dashed line) and $2.0\epsilon$ (purple dot-dashed line); (c) $3.0\epsilon$ (cyan solid line) and $4.0\epsilon$ (orange dashed line).}
 \label{fig:GofR_PP}
\end{figure}

The correlation among monomer beads, quantified as the monomer-monomer radial distribution function, $g_{pp}(r)$, is shown in Fig.~\ref{fig:GofR_PP} at various values of $\epsilon_{ps}$. When $\epsilon_{ps}$ is small, e.g., $\epsilon_{ps}=0.4\epsilon$ and $0.7\epsilon$, $g_{pp}(r)$ exhibits a huge first peak at $r\simeq 1\sigma$ and a small second peak at $r \simeq 2\sigma$, and then decays gradually toward $g_{pp}(r)=1$ at large $r$ ($> 10\sigma$). Subsequent peaks are barely visible. These features reflect the structure of dense globules formed by the polymer chains, with the average globule size and globule-globule separation controlling the characteristic length scales of $g_{pp}(r)$.

The $g_{pp}(r)$ curves for $\epsilon_{ps} = 0.95\epsilon$, $1.0\epsilon$, and $2.0\epsilon$ shown in Fig.~\ref{fig:GofR_PP}(b) are qualitatively different from those in Fig.~\ref{fig:GofR_PP}(a). The first peak at $r\simeq 1\sigma$ is still high as it contains the contribution from pairs of monomers in direct contact or bonded together. This first peak barely changes even if bonded pairs are excluded in the calculation of $g_{pp}(r)$. The second peak in $g_{pp}(r)$ at $r\simeq 2\sigma$ is much lower and the following decay is much quicker than those in the poor-solvent cases with small values of $\epsilon_{ps}$ shown in Fig.~\ref{fig:GofR_PP}(a), reflecting the extended chain conformations and the more uniform dispersion states of chains at these intermediate values of $\epsilon_{ps}$. Furthermore, the second peak is lower and the subsequent decay is more slowly as $\epsilon_{ps}$ is increased from $0.95\epsilon$ to $2.0\epsilon$. This is consistent with the early observation with a single-chain that the chain is more expanded at $\epsilon_{ps} = 2.0\epsilon$. Another feature reflected by the $g_{pp}(r)$ curve at $\epsilon_{ps} = 2.0\epsilon$ is the emergence of the deep trough around $r \simeq 1.4\sigma$, at which $g_{pp}(r) < 2$, between the first and second peaks. As discussed in more details below, this is due to the presence of solvent atoms in the gap between two monomers spatially near each other when the monomer-solvent interaction is strongly attractive, which makes the separation between a pair of closely distributed but not directly bonded monomers more likely at around $2\sigma$, the location of the second peak in $g_{pp}(r)$. As a result, there is a deficiency of monomer pairs separated by about $1.4\sigma$, causing the trough in $g_{pp}(r)$ at that location.

The radial distribution functions shown in Fig.~\ref{fig:GofR_PP}(c) for large values of $\epsilon_{ps}$ have several interesting features. After the high first peak at $r \simeq 1\sigma$, there is a deep trough at $r$ around $1.25\sigma$ where $g_{pp}(r)$ is about 0.5 at $\epsilon_{ps} = 3.0\epsilon$ and close to 0 at $\epsilon_{ps} = 4.0 \epsilon$. That is, there is a deficiency of monomer-monomer pairs at separations around $1.25\sigma$ and this deficiency becomes more dramatic as $\epsilon_{ps}$ is increased. The second peak in $g_{pp}(r)$ again occurs at $r\simeq 2\sigma$, corresponding to a pair of monomer beads bridged by a solvent atom. The subsequent peaks in $g_{pp}(r)$ correspond to more layers of solvent atoms in the gap between monomer beads. The declining behavior of $g_{pp}(r)$ at $r\gtrsim 2\sigma$ for $\epsilon_{ps} = 3.0\epsilon$ and $4.0\epsilon$ is similar to the cases with small values of $\epsilon_{ps}$ shown in Fig.~\ref{fig:GofR_PP}(a) but is even slower and more gradual. This behavior is a reflection of the aggregation state of polymer chains at strong monomer-solvent attractions. Eventually, at $r \simeq 50\sigma$, the $g_{pp}(r)$ curves for $\epsilon_{ps} = 3.0\epsilon$ and $4.0\epsilon$ decay to 1.

\begin{figure}[htb]
   \centering
   \includegraphics[width=0.4\textwidth]{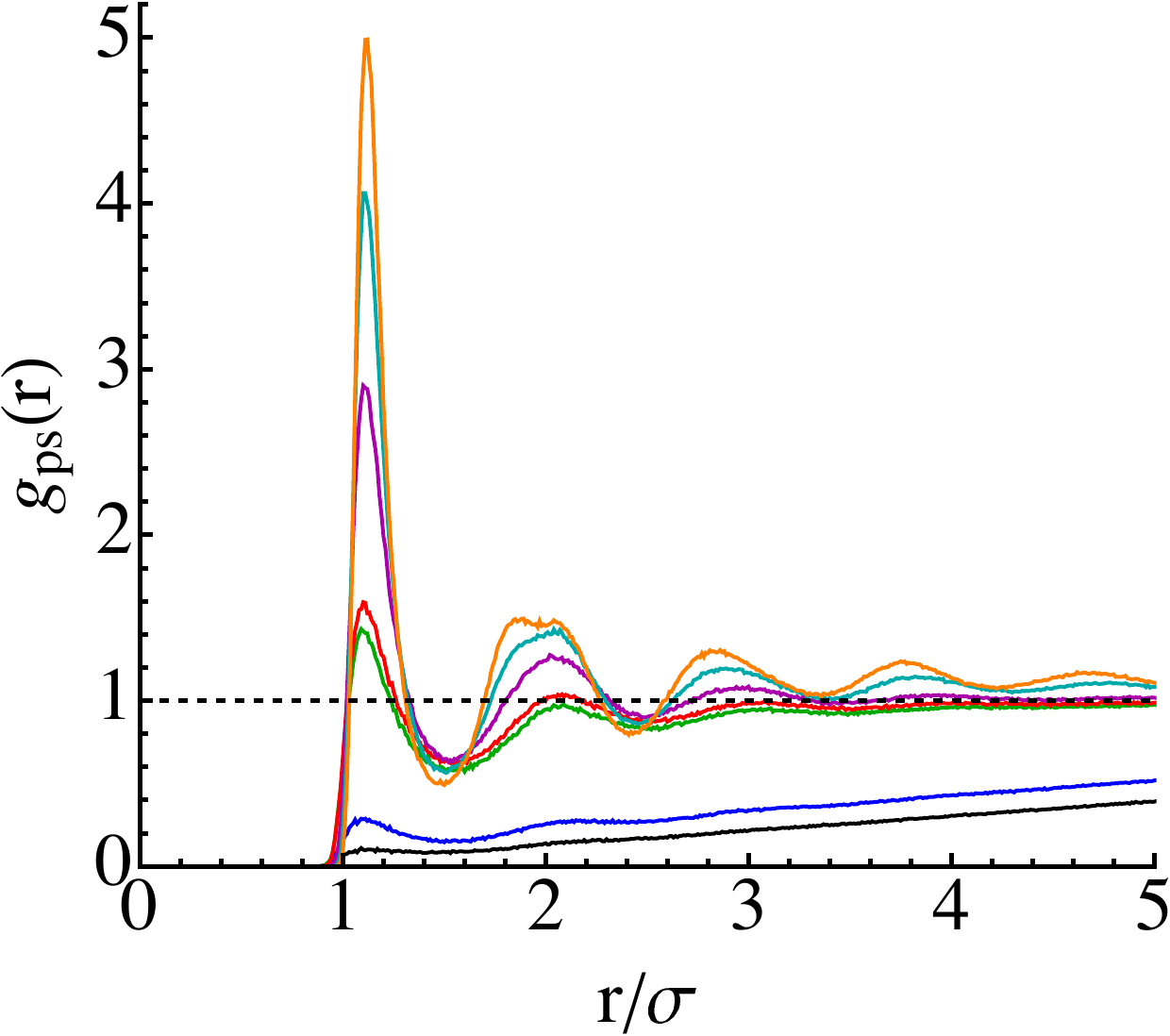}
   \caption{Monomer-solvent radial distribution function, $g_{ps}(r)$, at various values of $\epsilon_{ps}$: (a) $0.4\epsilon$ (black) and $0.7\epsilon$ (blue); (b) $0.95\epsilon$ (green), $1.0\epsilon$ (red) and $2.0\epsilon$ (purple); (c) $3.0\epsilon$ (cyan) and $4.0\epsilon$ (orange).}
 \label{fig:GofR_PS}
\end{figure}

The correlation among the distributions of monomers and solvent beads in the limit of small and large values of $\epsilon_{ps}$ revealed in the pair correlation function $g_{pp}(r)$ is directly revealed by the monomer-solvent radial distribution function, $g_{ps}(r)$. The results for various values of $\epsilon_{ps}$ are shown in Fig.~\ref{fig:GofR_PS}. At $\epsilon_{ps}= 0.4\epsilon$ and $0.7\epsilon$, the polymer chains phase separate from the solvent and the monomer beads are more likely in contact with each other. The first peak in $g_{ps}(r)$ therefore has a very small magnitude and then $g_{ps}(r)$ grows slowly, in a weak oscillatory manner, toward $g_{ps}(r)=1$ at large $r$. As $\epsilon_{ps}$ is increased, the solvent atoms are increasingly accumulated around the monomer beads, which is reflected by the monotonically increasing heights of the peaks and enhanced oscillations in $g_{ps}(r)$. At large values of $\epsilon_{ps}$, it is clear from $g_{ps}(r)$ that each monomer bead is surrounded and coated by a layer of solvent atoms.

\begin{figure}[htb]
   \centering
   \includegraphics[width=0.4\textwidth]{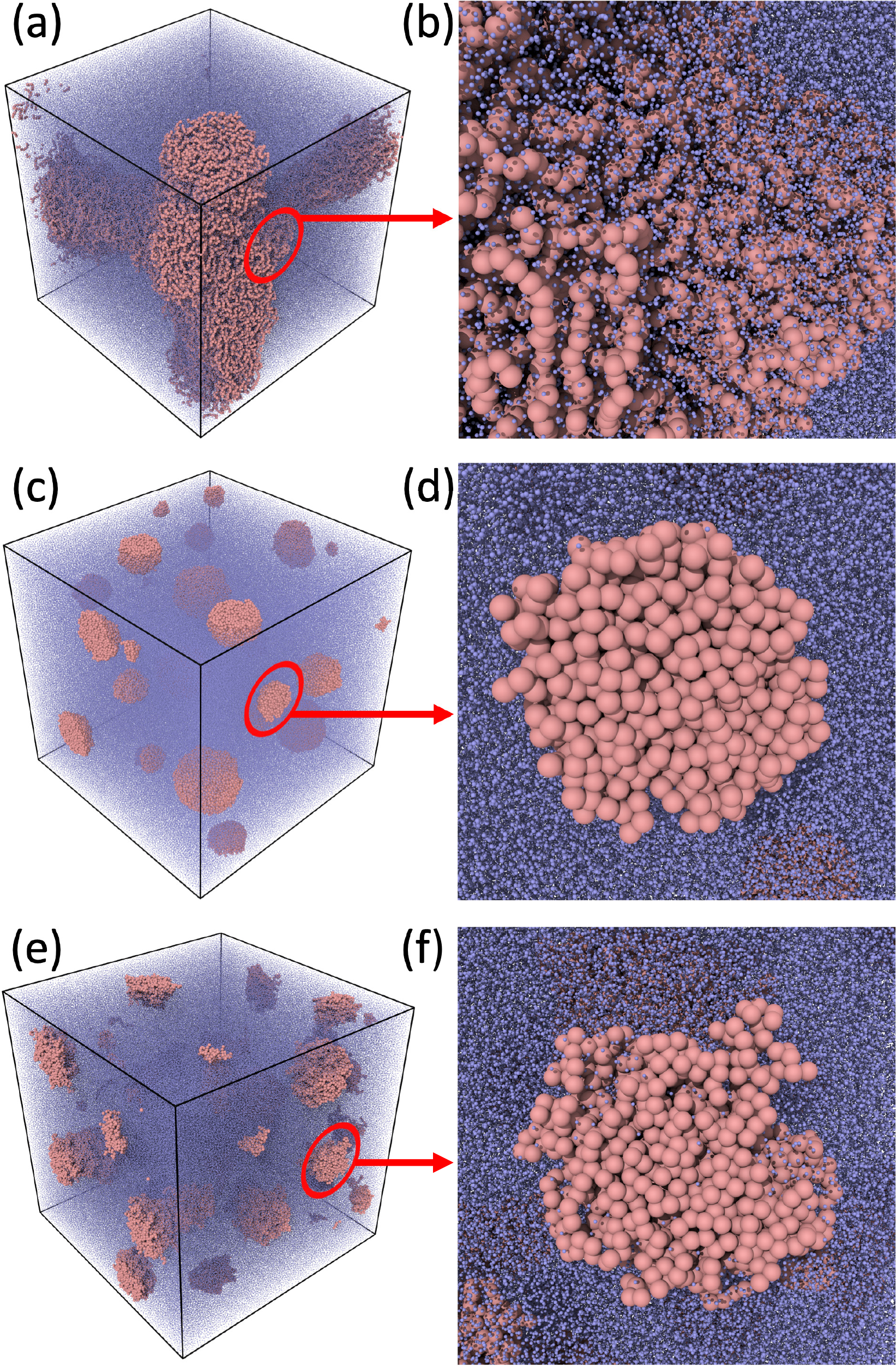}
   \caption{(a) Aggregation of 50-bead polymer chains and (b) a magnified view of a local region in the polymer aggregate at $\epsilon_{ps} = 3.0\epsilon$. (c) Globules of 50-bead polymer chains and (d) a magnified view of a region surrounding a globule at $\epsilon_{ps} = 0.4\epsilon$. (e) Globules of 50-bead polymer chains and (f) a magnified view of a region surrounding a globule at $\epsilon_{ps} = 0.9\epsilon$. For clarity, the monomer beads and solvent atoms are shown with a size ratio of 5:1.}
 \label{fig:eps_small_large}
\end{figure}

The features encoded in $g_{pp}(r)$ and $g_{ps}(r)$ can be corroborated with direct visualization of the polymer-rich domains, which is shown in Fig.~\ref{fig:eps_small_large}. At $\epsilon_{ps} = 3.0\epsilon$, the polymer chains form a network-like aggregate percolating the simulation box as shown in Fig.~\ref{fig:eps_small_large}(a). A snapshot of a small region in the aggregate is shown in Fig.~\ref{fig:eps_small_large}(b), where many solvent atoms are clearly visible in the gap between monomers. These solvent atoms have strong interactions with monomer beads and act like a glue to agglutinate chains together, providing the driving force for the chains to cluster and phase separate from the surrounding solvent.

The situation is quite different in the limit of weak monomer-solvent interactions (i.e., small values of $\epsilon_{ps}$). In the snapshots shown in Figs.~\ref{fig:eps_small_large}(c) and (d) for $\epsilon_{ps} = 0.4\epsilon$, the polymer chains form globules and in each globule, there is no solvent at all. That is, each globule consists purely of collapsed polymers. Therefore, although phase separation occurs at both small and large values of $\epsilon_{ps}$, the mechanism is quite different in the two limits. With weak monomer-solvent interactions, the solvent is quite poor and the polymer chains are fully collapsed as globules with a sharp monomer-solvent interface. However, in the limit of strong monomer-solvent attractions, the solvent plays the role of glue with respect to the polymer, causing the chains to partially collapse and the emergence of a polymer-rich aggregate of which the solvent is an integrated part. It remains an interesting question if this phase separation mechanism mediated by the solvent that interacts strongly with the polymer chains can be realized experimentally or is actually at play in certain scenarios. The key requirement is to enable some specific interactions between the monomers and solvent atoms/molecules to render the monomer-solvent interaction highly favorable and attractive.

The snapshots shown in Figs.~\ref{fig:eps_small_large}(e) and (f) are for $\epsilon_{ps} = 0.9\epsilon$, where the solvent is still poor for the chains with $N_m = 50$ but better than a nonsolvent. The chains still form globules. However, in each globule, the polymer chains are only partially collapsed and some solvent atoms are still present. This can be clearly seen in the magnified view of the region surrounding a part of a globule shown in Fig.~\ref{fig:eps_small_large}(f). However, the chains are more collapsed in these globules than those in the more extended aggregate that is formed at strong monomer-solvent attractions, such as the example shown in Fig.~\ref{fig:eps_small_large}(a) for $\epsilon_{ps} = 3.0\epsilon$.

\begin{figure}[htb]
   \centering
   \includegraphics[width=0.4\textwidth]{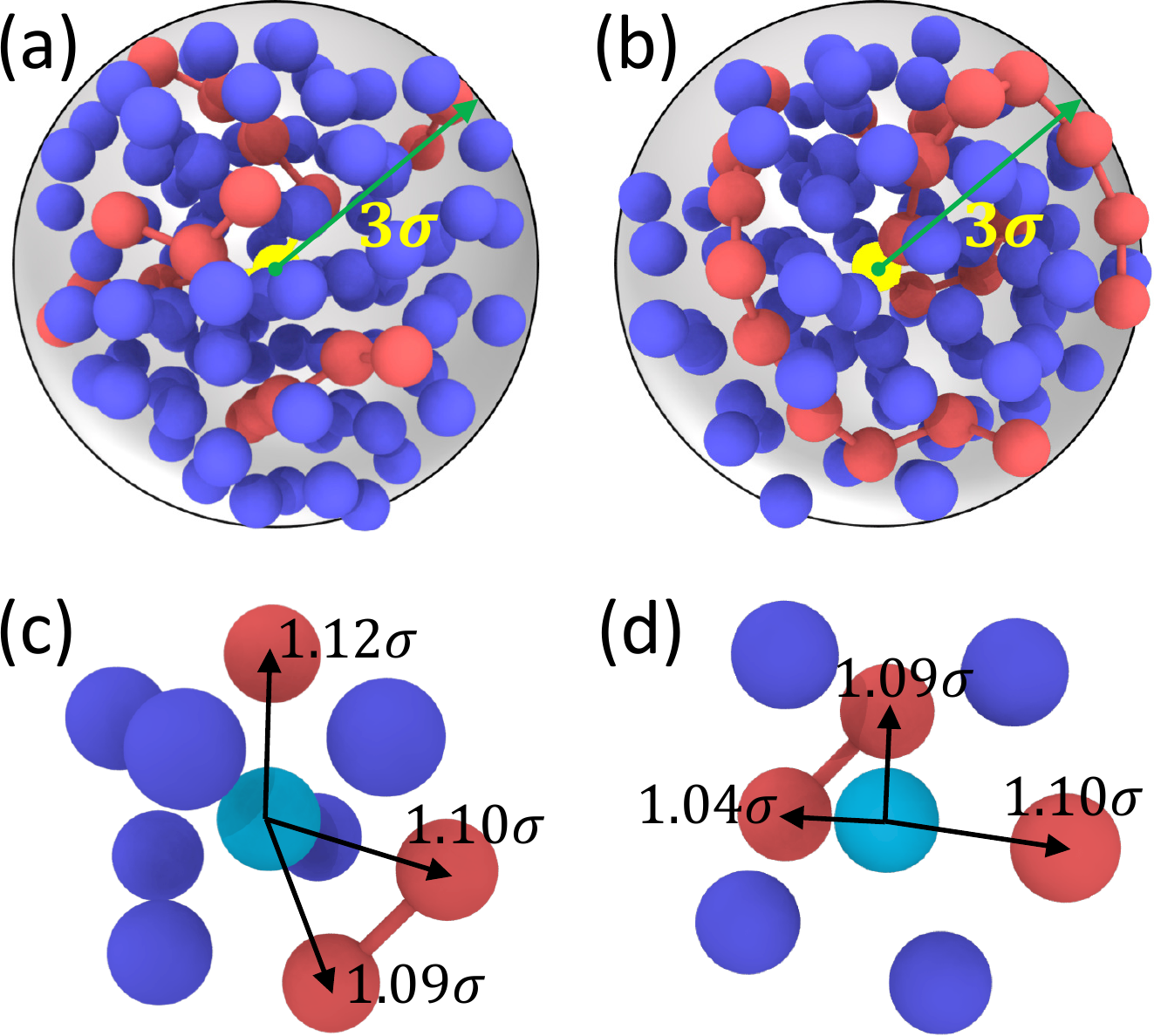}
   \caption{Solvent atoms (blue) and monomer beads (red) within $3\sigma$ from a chosen monomer (yellow) are imaged in (a) and (b) and those in contact with a chosen solvent atom (cyan) are imaged in (c) and (d) with the solvent-monomer separations labeled. The results are for the polymer solution [(a) and (c)] and the single-chain system [(b) and (d)] with $N_m=100$ at $\epsilon_{ps} = 4.0\epsilon$.}
 \label{fig:PS_contact}
\end{figure}

To more clearly illustrate the gluing effect of solvent atoms on monomers in the limit of strong polymer-solvent attractions, we show the snapshots of all solvent atoms and monomers within $3\sigma$, the cutoff of the LJ potentials, from a chosen monomer for the polymer solution in Fig.~\ref{fig:PS_contact}(a) and the single-chain system in Fig.~\ref{fig:PS_contact}(b) for $N_m=100$ at $\epsilon_{ps} = 4.0\epsilon$. The results show that the chains are partially collapsed in both systems in this limit and there are many solvent atoms and monomers within the interaction range of a given monomer. In Figs.~\ref{fig:PS_contact}(c) and (d), we further image monomers and solvent atoms in contact with a chosen solvent atom in the same two systems. Two particles are defined as contacting when their center-to-center separation is less than $2^{1/6}\sigma$, the location of the minimum of the LJ potential in Eq.~(\ref{eq:LJ}). This definition, frequently adopted in the contact mechanics literature\cite{Cheng2010TL}, suggests that two particles are in contact if their mutual force is repulsive. From the snapshots it is clear that a solvent atom is simultaneously in contact with multiple monomers from different chains (Fig.~\ref{fig:PS_contact}(c)) or segments of the same chain (Fig.~\ref{fig:PS_contact}(d)) as well as several solvent atoms. These results demonstrate that the solvent indeed glues polymers chains and force them to collapse when the polymer-solvent interactions are extremely attractive.

\section{DISCUSSION}

With MD simulations, an unexpected poor-solvent regime is discovered at strong polymer-solvent attractions. In the MD model employed here, the monomer-monomer and solvent-solvent interactions are kept identical with a fixed strength to provide a uniform reference as the monomer-solvent interaction strength is varied. For van der Waals interactions, the typical combination rule yields an inter-species interaction strength intermediate between the two intra-species strengths.\cite{Halgren1992} This indicates that the parameter $\lambda$, which characterizes the relative strength of the monomer-solvent interaction with respect to that of the solvent-solvent and nonbonded monomer-monomer interactions, should be about 1. However, in order to investigate a broad range of solvent qualities, we vary $\lambda$ from 0.4 to 4.0 in this work. To realize strong monomer-solvent attractions, some extra mechanisms are therefore needed to render the monomer-solvent interaction very favorable. Possible candidates include electrostatic interactions, dipole-dipole interactions, and hydrogen bonding, which are typically stronger than van der Waals interactions.\cite{Israelachvili2011} Experimentally, it will be a challenge to realize very strong polymer-solvent attractions. We expect that the computer simulation discovery of the poor-solvent behavior in this limit can motivate research in this direction.

To further simplify the MD model and reduce the number of tuning parameters, we adopt the same LJ length, $\sigma$, for both solvent atoms and monomer beads, which therefore have the same size. Furthermore, the LJ potentials for the solvent-solvent and nonbonded monomer-monomer interactions have the same strength. It is interesting to investigate the situations where Kuhn monomers are larger than solvent molecules and/or the polymer and solvent have different intra-species interaction strengths. Such scenarios will be explored in the future.

In the simulations reported in this paper, the temperature is fixed while the solvent quality is varied by changing the polymer-solvent interaction strength. The interesting result is the reentrance into a phase separation regime at very strong polymer-solvent attractions, in addition to the poor solvent regime expected at weak polymer-solvent attractions. This is qualitatively similar to systems possessing both lower (LCST) and upper critical solution temperatures (UCST) and with the LCST higher than the UCST, as discussed in the work of Clark and Lipson on polymer solutions and blends using an analytical lattice theory.\cite{Clark2012} In those systems, the solvent quality changes with temperature and phase separation occurs at both low and high temperatures, with a good solution only appearing in a window of intermediate temperatures. An important feature is that the enthalpy of mixing is negative at the LCST but turns to positive at the UCST.\cite{Clark2012} However, it is unclear if the phase separation mechanism at temperatures higher than the LCST is the same as that at temperatures lower than the UCST. In the work reported here, the reentrance transition is driven by the polymer-solvent interaction only at a constant temperature and the mechanism of phase separation at strong polymer-solvent attractions clearly differs from the one at weak attractions.

\section{CONCLUSIONS}

Molecular dynamics (MD) simulations are employed to study polymer solutions. In our simulations, Lennard-Jones interactions with an attractive tail are used for all the nonbonded interactions. The identical solvent-solvent and nonbonded monomer-monomer interactions are unchanged and kept as a reference. The strength of the monomer-solvent interaction is varied from weak to strong, in order to vary the solvent quality. We have uncovered an interesting chain-collapsing behavior and phase-separation mechanism in polymer solutions with strong attractions between monomers and solvent atoms. In this case, the solvent acts as a glue to adhere monomers together, causing a polymer chain to collapse and multiple chains to aggregate. As a result, two phases emerge in the polymer solution, one of which is the pure solvent while the other is an extended agglomeration of the polymer chains and solvent. This mechanism of phase separation is different from the case with weak monomer-solvent attractions, where the polymer chains are collapsed and form globules. In each globule, monomers contact each other and the solvent is either completely excluded from it (for a nonsolvent) or rather sparsely distributed (for a poor solvent).

With either weak or very strong monomer-solvent attractions, the Flory exponent is found to be around 0.3, indicating a nonsolvent or poor-solvent behavior. In the range of intermediate monomer-solvent attractions, the solvent is good with the Flory exponent around 0.6. Therefore there are two $\theta$-transitions as the strength of the monomer-solvent attraction is varied, one at a strength about 95\% and another between 190\% and 200\% of that of the solvent-solvent and nonbonded monomer-monomer interactions. In these $\theta$-solvents, the Flory exponent is about 0.5. Our results on the Flory exponent in solvents of various qualities are close to the predictions based on integral equations derived from the polymer Kirkwood hierarchy.\cite{Gan1998}

\subsection*{ACKNOWLEDGMENTS}

This material is based upon work supported by the National Science Foundation under Grant No. DMR-1944887. This research used resources of the National Energy Research Scientific Computing Center (NERSC), a U.S. Department of Energy Office of Science User Facility operated under Contract No. DE-AC02-05CH11231. These resources were obtained through the Advanced Scientific Computing Research (ASCR) Leadership Computing Challenge (ALCC). The authors thank Dr. Gary S. Grest for constructive criticism of the manuscript. The authors acknowledge Advanced Research Computing at Virginia Tech (URL: http://www.arc.vt.edu) for providing computational resources and technical support that have contributed to the results reported within this paper. S.C. gratefully acknowledges the support of NVIDIA Corporation with the donation of the Tesla K40 GPUs used for this research.




\clearpage
\newpage
\onecolumngrid
\renewcommand{\thefigure}{S\arabic{figure}}
\setcounter{figure}{0}
\renewcommand{\theequation}{S\arabic{equation}}
\setcounter{equation}{0} 
\renewcommand{\thepage}{SI-\arabic{page}}
\setcounter{page}{1}    
\begin{center}
{\bf SUPPORTING INFORMATION}
\end{center}

\noindent \textbf{S1. Additional Results}

\begin{figure}[H]
   \centering
   \includegraphics[width=0.5\textwidth]{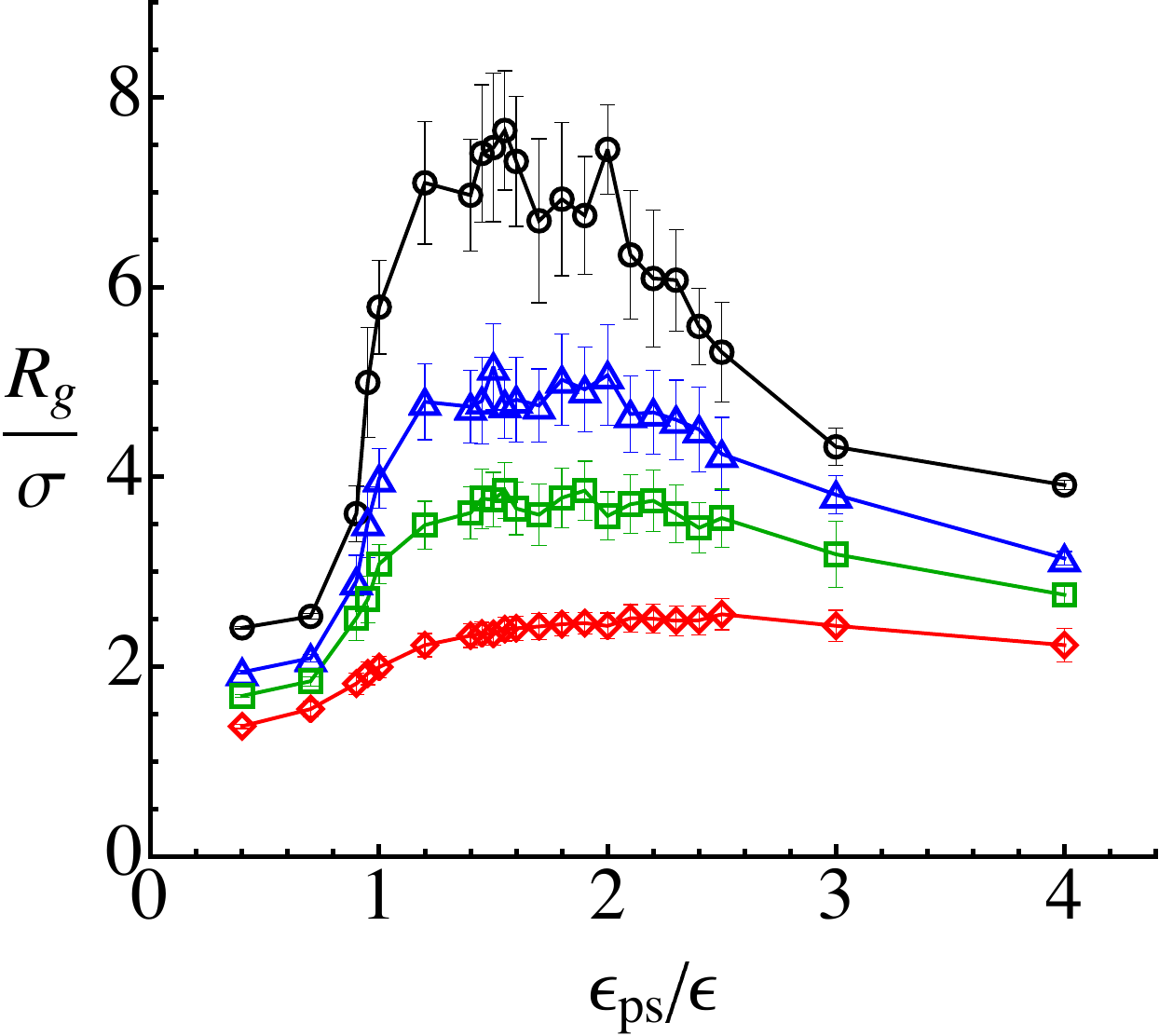}
   \caption{Radius of gyration ($R_g$) vs. monomer-solvent interaction strength ($\epsilon_{ps}$) for $N_m=16$ (diamonds), $32$ (squares), $50$ (triangles), and $100$ (circles).}
 \label{fig:Rg_vs_Eps_4N}
\end{figure}

\noindent Here in Fig.~\ref{fig:Rg_vs_Eps_4N} we include additional results on the radius of gyration ($R_g$) plotted against the monomer-solvent interaction strength ($\epsilon_{ps}$) not shown in the main text.


\begin{figure}[htb]
   \centering
   \includegraphics[width=\textwidth]{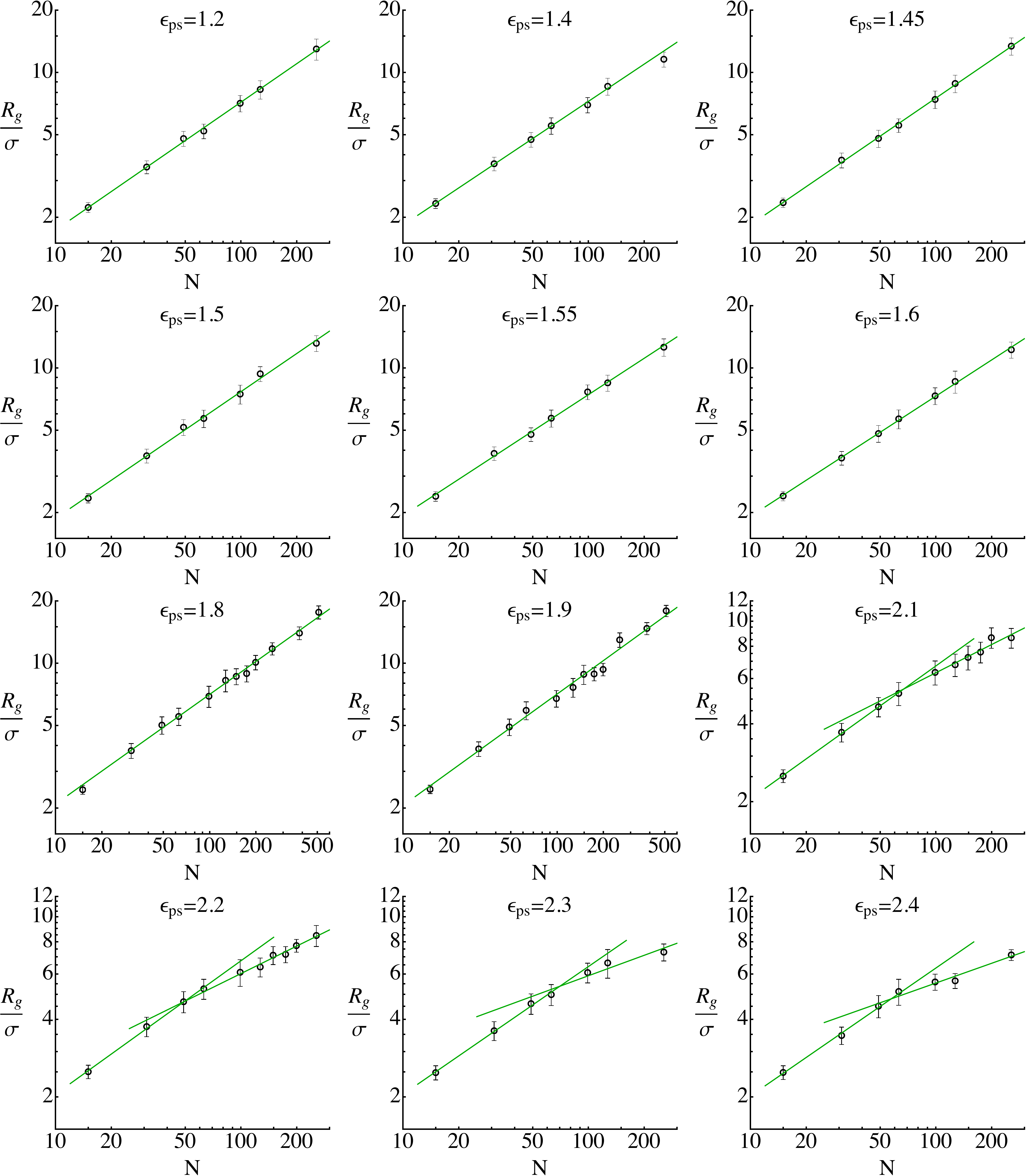}
   \caption{Radius of gyration ($R_g$) vs. number of bonds ($N$) at various values of $\epsilon_{ps}$.}
 \label{fig:Rg_vs_N_other}
\end{figure}

\noindent In Fig.~\ref{fig:Rg_vs_N_other} the results on $R_g$ vs. the chain length ($N$) at various values of $\epsilon_{ps}$ not plotted in the main text are shown.

\begin{figure}[htb]
   \centering
   \includegraphics[width=\textwidth]{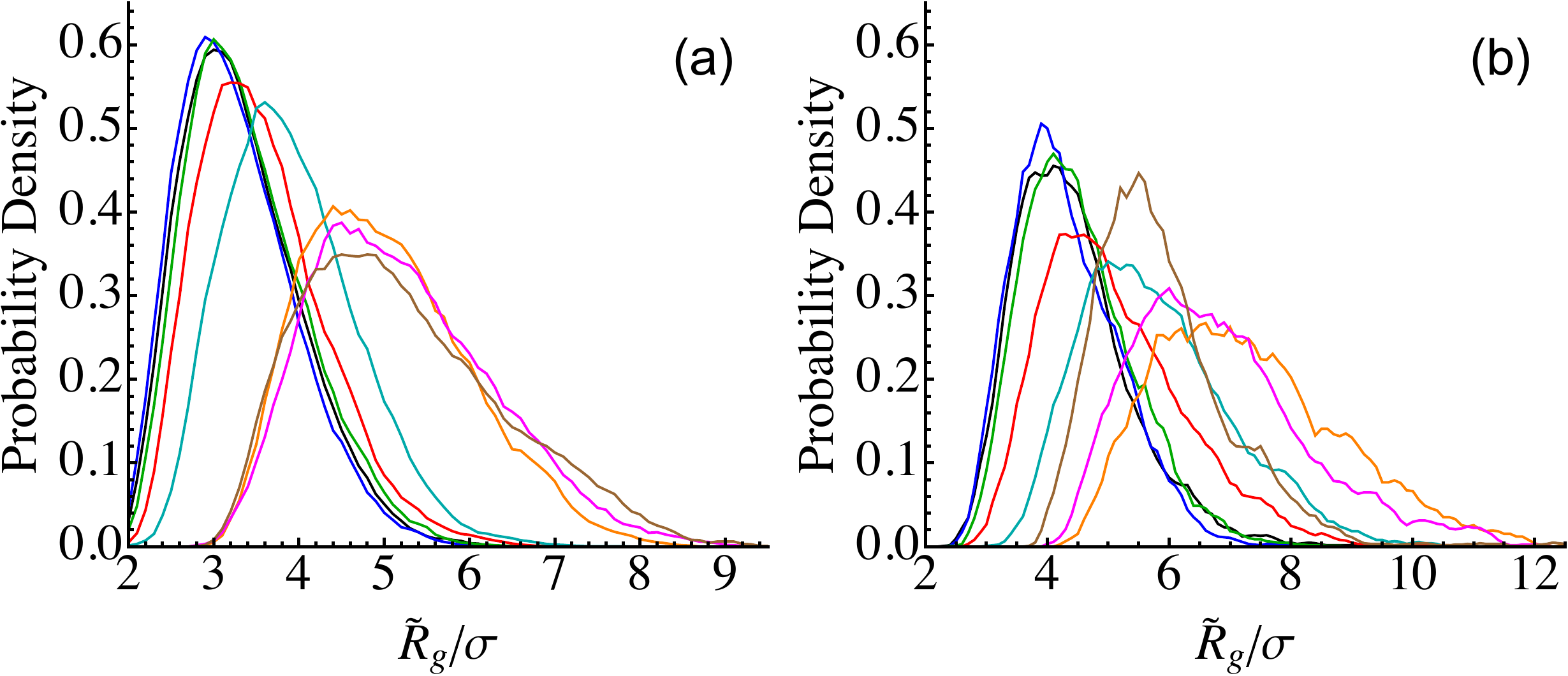}
   \caption{Probability distribution of the instantaneous values of the radius of gyration ($\tilde{R}_g$) at $\epsilon_{ps}/\epsilon = 0.4$ (black), 0.7 (blue), 0.9 (green), 0.95 (red), 1.0 (cyan), 2.0 (orange), 3.0 (magenta), and 4.0 (brown). The data are for the 1000 chains with $N_m = 50$ or the 500 chains with $N_m = 100$ in the polymer solutions as those shown in Figure 6 of the main text.}
 \label{fig:Rg_dist_polym_sol}
\end{figure}

\noindent in Fig.~\ref{fig:Rg_dist_polym_sol} the probability distribution of the instantaneous values of the radius of gyration is plotted. The data are collected from the polymer solutions containing either 1000 chains with $N_m = 50$ or 500 chains with $N_m = 100$. The distribution is unimodal, indicating that all the chains in a given polymer solution behave similarly in a statistical sense.

\end{document}